\documentclass[10pt]{article}
\usepackage[letterpaper]{geometry}
\usepackage{hicss}
\usepackage{times}
\usepackage{hyphenat}
\usepackage[hidelinks]{hyperref}
\usepackage{url}
\usepackage{latexsym}
\usepackage{indentfirst}
\usepackage{booktabs}
\usepackage{siunitx}
\usepackage{threeparttable}
\usepackage[
backend=biber,
style=apa,
natbib=true,
maxbibnames=99,
firstinits=true
]{biblatex}
\usepackage{graphicx}
\graphicspath{{Figures/}}
\title{Quantifying Visual Properties of GAM Shape Plots: Impact on Perceived Cognitive Load and Interpretability}

\author{
  \begin{tabular}{@{}c@{}}
    \begin{tabular}{ccc}
      Sven Kruschel & Lasse Bohlen & Julian Rosenberger \\
      University of Regensburg & Leipzig University & University of Regensburg \\
      \underline{sven.kruschel@ur.de} & \underline{lasse.bohlen@uni-leipzig.de} & \underline{julian.rosenberger@ur.de}
    \end{tabular} \\[2em]
    \begin{tabular}{cc}
      Patrick Zschech & Mathias Kraus \\
      Leipzig University & University of Regensburg \\
      \underline{patrick.zschech@uni-leipzig.de} & \underline{mathias.kraus@ur.de}
    \end{tabular}
  \end{tabular}
}

\begin{document}
\maketitle
\begin{abstract}
Generalized Additive Models (GAMs) offer a balance between performance and interpretability in machine learning. The interpretability aspect of GAMs is expressed through shape plots, representing the model's decision-making process. However, the visual properties of these plots, e.g. number of kinks (number of local maxima and minima), can impact their complexity and the cognitive load imposed on the viewer, compromising interpretability. Our study, including 57 participants, investigates the relationship between the visual properties of GAM shape plots and cognitive load they induce. We quantify various visual properties of shape plots and evaluate their alignment with participants' perceived cognitive load, based on 144 plots. Our results indicate that the number of kinks metric is the most effective, explaining 86.4\% of the variance in users' ratings. We develop a simple model based on number of kinks that provides a practical tool for predicting cognitive load, enabling the assessment of one aspect of GAM interpretability without direct user involvement.
\end{abstract}

\subsubsection*{Keywords:}

Interpretable Machine Learning, Generalized Additive Models, Shape Plots, Visual Properties, Cognitive Load

\section{Introduction}
\label{chap:intro}

Machine learning (ML) has become ubiquitous, transforming data analysis, prediction, and decision-making across various domains, such as sales, healthcare, and finance \citep{janiesch2021machine}.
%from low-stakes applications like sales prediction \citep{chen2023attending} to high-stakes fields such as healthcare \citep{esteva2017dermatologist}, finance \citep{ban2018machine}, and justice \citep{thiebes2021trustworthy}.
As ML models become increasingly prevalent, particularly in critical applications, the need for their interpretability has become a pressing concern \citep{meske2022explainable}.

Generalized Additive Models (GAMs) have emerged as a promising approach for interpretable ML models, aiming to balance performance and interpretability \citep{zschech2022gam}. GAMs are based on the additive combination of shape functions, which represent the relationship between individual input features and the model's output \citep{lou2012intelligible}. This additive structure allows GAMs to be represented by shape plots, providing a visual representation of the model's decision logic and thus enabling model interpretation \citep{chang2021interpretable}.
% However, the interpretability of ML models, including GAMs, is a complex and multifaceted concept that lacks a clear definition and quantification \citep{lipton2018mythos, nauta2023anecdotal, doshi2017towards}. This ambiguity is intensified by the fact that interpretability is often subjective and dependent on the diverse needs and backgrounds of various stakeholders \citep{hohman2019gamut, tomsett2018whom, forster2023user}. We argue that interpretability cannot be directly measured or manipulated as a single monolithic property \citep{lipton2018mythos}. Therefore, we follow \citet{poursabzi2021manipulating} in viewing it as a human-centric concept influenced by various \textit{manipulable factors} and assessed through \textit{measurable (human) outcomes}.
However, interpretability is a complex and multifaceted concept that lacks a clear definition and quantification \citep{lipton2018mythos, nauta2023anecdotal, doshi2017towards}. We argue that interpretability is a human-centric concept influenced by various \textit{manipulable factors} and assessed through \textit{measurable (human) outcomes} \citep{poursabzi2021manipulating}.

In the context of GAMs, the visual properties of shape plots, such as graph length and number of kinks (number of local maxima and minima), can vary significantly depending on the model architecture and hyperparameters \citep{chang2021interpretable}. These visual properties are hypothesized to influence the cognitive load imposed on the viewer, potentially affecting their ability to understand and interpret a model's decision-making process \citep{abdul2020cogam, shah2002review}.

This study explores the relationship between the visual properties of GAM shape plots and their associated cognitive load. We quantify the visual properties using Python functions that generate numeric metrics, enabling efficient and reproducible analysis. Through a user study, we evaluate the alignment between these metrics and participants' perceived cognitive load. Our findings indicate that the number of kinks metric is the most effective among those tested, accounting for 86.4\% of the variance in user ratings and serving as a precise predictor of cognitive load in GAM shape plots.
By identifying the visual properties that most influence cognitive load and providing Python functions to extract them, our findings can guide the design of more interpretable GAMs. We propose objective metrics that quantify these visual properties and assess their alignment with user perceptions of cognitive load. Our context-independent approach allows to evaluate one facet of GAM interpretability based on shape plots. This enables comparison and refinement of GAMs without direct user involvement. 
Our research contributes to the literature by proposing a novel approach to quantify the visual properties of GAM shape plots and identifying the number of kinks metric as the most effective predictor of cognitive load among the tested. We also provide access to a public dataset of shape plots with user-rated cognitive load, facilitating future research on assessing the interpretability of GAM shape plots.\footnote{The dataset and supplementary materials are available at: \url{https://osf.io/ucjtq/?view_only=bdb6b9d5d7994fa79ad0fd9826b44e61}}

The remainder of this paper is structured as follows. Section~\ref{sec:related_work} presents the related work, while Section~\ref{sec:methods} elaborates on our applied methodology. Section~\ref{chap:results} demonstrates the effectiveness of our approach in quantifying cognitive load in GAM shape plots. Lastly, Section~\ref{chap:discussion} discusses the limitations of our work as well as the implications for future development of interpretable ML models and concludes our work.

% % ========================
% % Chapter: Related Work (3.5 pages)
% % ========================

\section{Related Work}
\label{sec:related_work}

\subsection{Interpretable Machine Learning}
\label{sec:interp_ml}

As ML models are increasingly deployed in critical domains such as healthcare and finance, the demand for interpretability has grown \citep{meske2022explainable}. Interpretable models provide insights into their decision-making process, enabling users to understand a model's prediction and thereby ensure consistency with domain knowledge \citep{doshi2017towards}.

However, interpretability is a complex and multi-faceted concept that lacks a clear definition and quantification \citep{nauta2023anecdotal}. Different notions of interpretability, such as simplicity, transparency, and simulatability are often conflated \citep{lipton2018mythos}. Moreover, interpretability needs may vary depending on the stakeholders and specific use cases \citep{forster2023user, tomsett2018whom}. To address these challenges, \citet{poursabzi2021manipulating} propose a framework that conceptualizes interpretability as a human-centric property influenced by \textit{manipulable factors} and captured through \textit{measurable (human) outcomes}. In the context of our study, the visual properties of GAM shape plots serve as manipulable factors that are hypothesized to influence cognitive load, a measurable human outcome. By investigating the relationship between these visual properties and cognitive load, we aim to provide insights into one specific aspect of GAM interpretability. This approach aligns with the framework proposed by \citet{poursabzi2021manipulating} and contributes to a better understanding of the factors that impact the interpretability of GAMs.

\subsection{Generalized Additive Models (GAMs)}
\label{sec:gams}

GAMs have emerged as a promising approach to interpretable ML, offering a balance between performance and interpretability \citep{zschech2022gam, kraus2024interpretable}. GAMs represent the relationship between input features and the target variable using univariate functions, often referred to as shape functions \citep{lou2012intelligible}. The final prediction is obtained by summing the contributions of each shape function:

\vspace{-0.4cm}

\begin{equation}
f(x) = f_1(x_1) + f_2(x_2) + \ldots + f_n(x_n),
\label{eq:gam_prediction}
\end{equation}
where each $f_i$ denotes a shape function of the $i$-th input feature $x_i$ to the prediction target. 

\begin{figure*}[ht]
\centering
\includegraphics[width=0.7\textwidth]{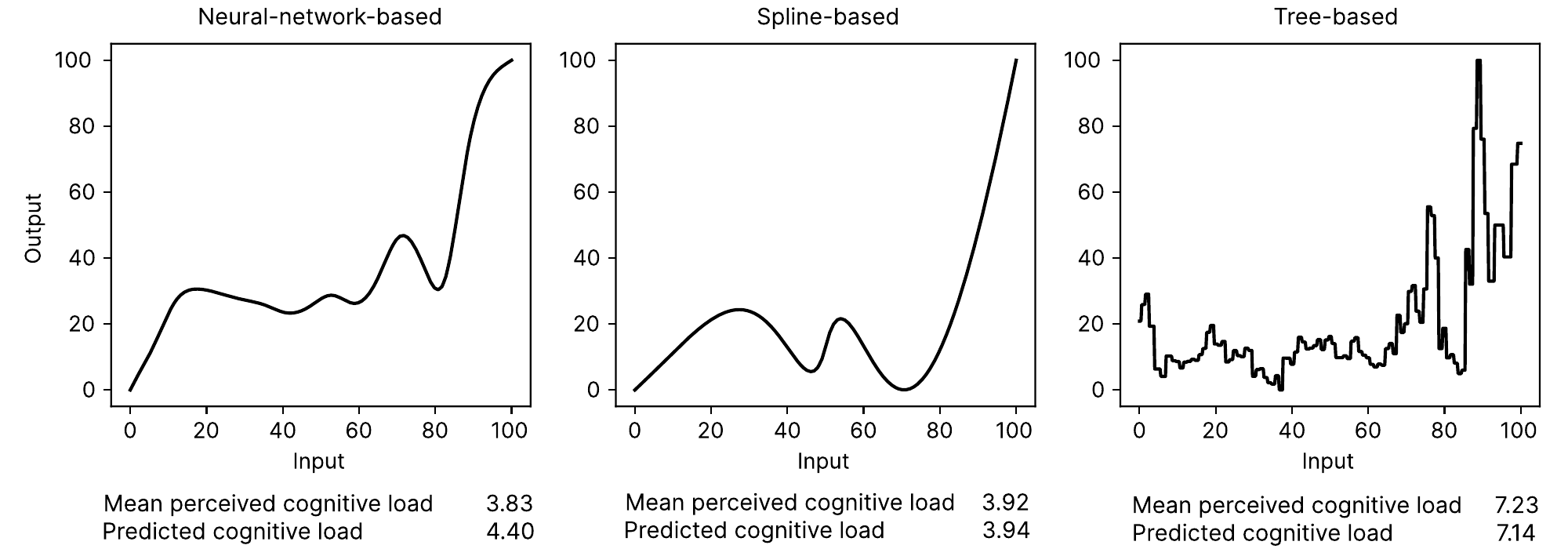}
\caption{Visual properties of shape plots for the same feature vary based on underlying model architecture, thereby affecting both users' perceived cognitive load and predicted cognitive load.}
\label{fig:plot-overview}
\end{figure*}

One of the key advantages of GAMs is their interpretability, due to their additive structure. GAMs allow for the visualization of individual feature contributions through shape plots. These plots display the relationship between each feature and the target variable, holding all other features constant. By examining shape plots, users can gain insights into how each feature affects the model's predictions and identify potential flaws \citep{chang2021interpretable}.
However, the visual properties of shape plots can vary significantly depending on the model architecture and hyperparameters. This results in shape plots ranging from smooth curves to step plots, as shown in Figure~\ref{fig:plot-overview}. These variations in visual properties can impact the interpretability of GAMs, as some shape plots may be more difficult to interpret than others \citep{abdul2020cogam, chang2021interpretable}. 

% One of the key advantages of GAMs is their interpretability, which stems from their additive structure. GAMs allow for the visualization of individual feature contributions through shape plots \citep{chang2021interpretable}. These plots, exemplified in Figure~\ref{fig:plot-overview}, display the relationship between each feature and the target variable, holding all other features constant. By examining shape plots, users can gain insights into how each feature affects the model's predictions and identify potential non-linearities. However, the visual properties of shape plots can vary significantly depending on the model architecture and hyperparameter choices, as GAMs primarily learn shape functions based on patterns in the data rather than establishing causal relationships. This results in shape plots ranging from smooth curves to step plots, as shown in Figure~\ref{fig:plot-overview}. These plots show the input-output relationships for the same feature in the same dataset, as learned by three different GAMs. For example, in the plot on the right, an input value of 20 positively affects the overall ML model by 18, but accurately reading and interpreting values around the input value of 80 can be challenging compared to the smoother plots on the left.

% As discussed in the previous sections, interpretability is a multi-faceted concept influenced by various factors and manifested through measurable human outcomes \citep{poursabzi2021manipulating}. The effectiveness of shape plots in supporting human understanding and decision-making depends on their visual properties. In this study, we focus specifically on numerical shape plots, investigating how their visual properties impact cognitive load, a measurable human outcome that can impede interpretability.

\subsection{Visual Properties, Graph Complexity, and Cognitive Load}
\label{sec:vp_gc_cl}

%The interpretability of GAMs is closely tied to the visual properties of their shape plots. As discussed in Section~\ref{sec:interp_ml},
As introduced earlier, interpretability as a human-centered concept can be influenced by manipulable factors and assessed by measurable human outcomes \citep{poursabzi2021manipulating}. In the context of GAMs, manipulable factors are the visual properties of shape plots, such as graph length and visual chunks, while the cognitive load imposed by the shape plots can be seen as a human-centered, measurable aspect of interpretability \citep{abdul2020cogam}.

% Visual properties have been studied to assess graph complexity and research in cognitive psychology suggests that more complex graphs require more mental effort to understand \citep{carswell1993stimulus, shah2002review, peebles2003modeling}. As the visual properties of shape plots change, such as increased graph length or a higher number of kinks, the mental effort imposed on the viewer is expected to rise, potentially affecting the effort associated with processing the information.

Cognitive load refers to the mental effort required to process and understand information \citep{sweller1988cognitive}. In the context of graph comprehension, cognitive load is influenced by various factors, such as graph complexity \citep{huang2009measuring}, user expertise, and domain knowledge \citep{tomsett2018whom}. Higher cognitive load may hinder the effective interpretation of shape plots, thereby potentially reducing the interpretability of GAMs \citep{abdul2020cogam}. However, it is important to acknowledge that the relationship between cognitive load and interpretability is nuanced and varies with specific contexts.

Visual properties have been studied to assess graph complexity, and research in cognitive psychology suggests that more complex graphs require more mental effort to understand \citep{shah2002review, peebles2003modeling}. A common approach to quantify graph complexity is the use of visual chunks, which are generally defined as the areas between significant deviations in a graph's trend, such as trend reversals \citep{carswell1993stimulus, abdul2020cogam}. The underlying assumption is that a higher number of visual chunks indicates increased graph complexity, which in turn leads to higher cognitive load. However, we argue that visual chunks capture only one facet of graph complexity and may not be the most comprehensive metric for assessing the cognitive load imposed by a graph. Other visual properties, such as graph length and the number of kinks, might represent the overall complexity and the mental effort required to process the information similarly well.

% In the graph cognition literature, graph complexity is often quantified in terms of visual chunks, which are generally defined as the areas between two trend reversals in a graph. It is argued that more chunks increase graph complexity and thereby cognitive load. We argue that this more complex quantification captures only one aspect of graph complexity and may not be the best metric for assessing the cognitive load imposed by a graph.

Our study focuses on quantifying multiple visual properties of shape plots as objective metrics and modeling their impact on cognitive load. In addition to visual chunks, we consider other metrics, which are directly related to the visual properties discussed earlier (see Section \ref{sec:metrics_development}). By assessing the alignment between these metrics and user perceptions of cognitive load, we aim to validate them as measures of graph complexity in GAM shape plots. The proposed metrics offer a context-independent, numerical method for evaluating one dimension of GAM interpretability based on shape plots, enabling comparison and refinement of GAMs without direct user involvement. However, we acknowledge that our approach does not constitute a comprehensive interpretability assessment, as interpretability is a multi-faceted concept that extends beyond visual properties.

\section{Methodology}
\label{sec:methods}

%Our study investigates the interpretability of GAM shape plots by adopting a human-centric approach, where we quantify the visual properties of shape plots as manipulable factors and assess their alignment with user perceptions of cognitive load as a measurable human outcome \citep{poursabzi2021manipulating}. 

% First, we develop a set of metrics to quantify the visual properties of shape plots, capturing aspects such as graph length, number of kinks, and polynomial degree. Second, to validate these metrics, we conduct a user study using a diverse set of shape plots generated by applying three distinct GAMs to 20 real-world benchmark datasets. From the resulting pool of 1120 plots, we select 144 for our experiment. We divide the 144 plots into 9 groups á 16 plots. Each participant rates 2 plot groups, hence 32 plots for cognitive load, makes 8 binary decisions for another plot group, and ranks another group in 4 sets á 4 plots. This crossover approach allows us to draw more generalizable conclusions.
% To select the 144 plots, we sort the 1120 plots on each of 14 initial metric values, with 16 plots selected at equal intervals per metric. The resulting 224 plots are randomly assigned into the 9 groups to ensure an equal wide range of shapes and complexities in each group. The plot axes are standardized, and contextual information is removed to isolate the effect of visual properties on user perception.

We develop metrics to quantify visual properties of shape plots, such as graph length, number of kinks, and polynomial degree. To validate these, we conduct a user study using 144 plots (9 sets of 16 plots each, matching our 9 participant groups). Those are selected from a pool of 1120 plots, generated by applying three distinct types of GAMs to 20 real-world datasets. Our 57 participants, divided into 9 groups, complete three tasks with different plot sets: ranking 16 plots, making 8 decisions for binary choices on plot complexity with 16 different plots, and rating 32 plots (two sets of plots) on cognitive load. We use a rotating approach where each group interacts with different sets for each task. For example, Group 1 ranks Set 1, makes binary decisions on Set 2, and rates Sets 3 \& 4; Group 2 ranks Set 2, decides on Set 3, and rates Sets 4 \& 5, and so on. This ensures each plot is evaluated multiple times across different assessment types, allowing for more generalizable conclusions.

The actual plot selection involves sorting the 1120 plots on 14 initial metric values, selecting 16 plots at equal intervals per metric. The resulting 224 plots (14 × 16) are then randomly distributed into 9 sets of 16 plots each, yielding our final selection of 144 plots. Plot axes are standardized and contextual information removed to isolate visual property effects.

Participants rate perceived mental effort using the 9-point scale by \citet{paas1992training}, widely accepted in cognitive science. This data assesses cognitive load imposed by GAM shape plots, linking to interpretability. We use it to create linear models predicting cognitive load based on our visual property metrics. Further details on our methodology, including specific metrics and selection rationale, can be found in our supplementary materials.

\subsection{Visual Property Metrics Development}
\label{sec:metrics_development}

To evaluate all GAMs using the same set of metrics, we create Python functions to automatically extract visual properties from a normalized set of points of any shape plot. This ensures a uniform evaluation across all plots, regardless of their continuity or differentiability. Initially, we systematically decompose shape plots into characteristics like the depth of curves or number of inflection points. While these initial 25 metrics were not directly derived from existing literature, they were grounded in fundamental principles of graph analysis and visual complexity. Using a data-driven approach, we successively exclude redundant or non-informative metrics. Through a final analytical assessment, we correlate the remaining 14 metrics\footnote{Detailed descriptions of these metrics and the selection procedure are available in our supplementary materials: \url{https://osf.io/dc4e2?view_only=bdb6b9d5d7994fa79ad0fd9826b44e61}} with perceived complexity to select the most promising metrics. As a double-check, we recalculate discarded metrics post-experiment, confirming that our five metrics are relevant and effective while maintaining a manageable scope. Each final metric quantifies a different visual property of shape plots, capturing various aspects of graph complexity:

\textbf{Graph Length.} A graph's length provides a cumulative measure of its complexity, representing the total trajectory from start to end. A longer graph typically suggests greater complexity.

\textbf{Polynomial Degree.} This metric identifies the lowest polynomial degree that can accurately approximate the GAM shape curve with minimal error. A lower degree indicates a simpler curve, while a higher degree indicates greater complexity with more variations.

\textbf{Visual Chunks.} This metric focuses on the segmentation of the graph into distinct segments or chunks. In the graph cognition literature, a visual chunk in a line graph is generally defined as the area between two trend reversals \citep{carswell1993stimulus}. A trend reversal occurs when the direction of a trend changes, either from increasing to decreasing or vice versa. The number of visual chunks is calculated by counting the number of areas between all trend reversals in the shape plot. In line with research from cognitive psychology, we expect that graphs with more visual chunks are more complex and lead to higher cognitive load \citep{abdul2020cogam}.

\textbf{Number of Kinks.} This metric counts the sum of local maxima and minima within the shape plot, offering insight into its detailed characteristics. Formally, we define a maxima as a point where the function stops increasing and starts decreasing, and a minima vice versa. The underlying assumption is that a graph with more kinks is perceived as more complex, which results in higher cognitive load.

\textbf{Average Kink Distance.} Not only the number of kinks are relevant, but also the distance between each kink. This metric evaluates the mean distance between kinks, offering a measure of the graph's density and frequency of changes. A high average distance indicates a simpler graph with distinct kinks, while a lower average distance suggests a noisy graph with many kinks close to each other. For our analysis, we use the inverse measure, ensuring that as the metric increases, so does a shape plot's perceived complexity.

\begin{figure*}[h!]
\centering
\includegraphics[width=0.78\textwidth]{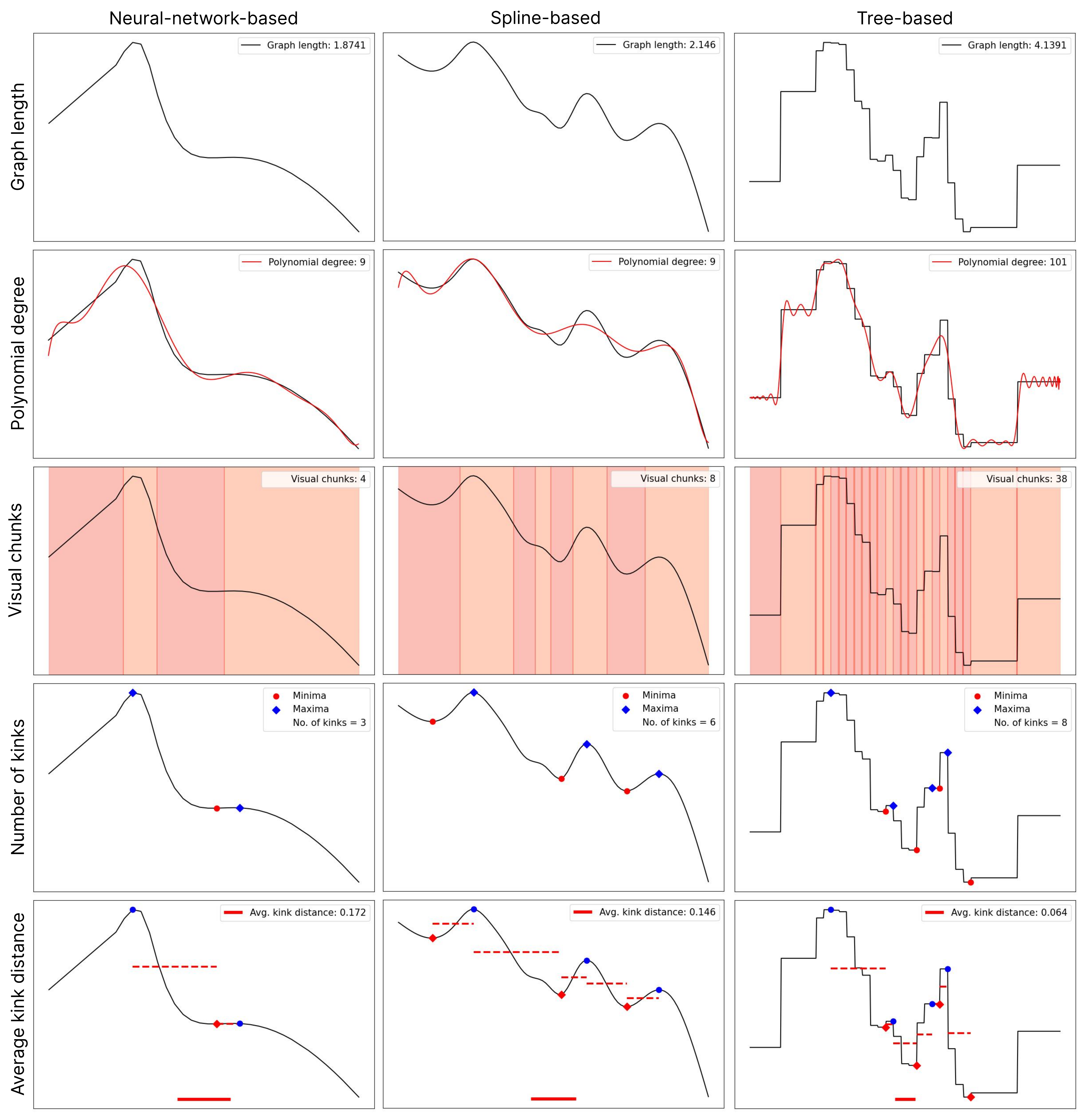}
\caption{Visual representation of the metrics for three different shape plots. The columns indicate three different types of GAMs. The rows represent the five metrics selected in several workshops.}
\label{fig:metrics}
\end{figure*}

Figure \ref{fig:metrics} visualizes the metrics for three different shape plots, each representing the same feature from the same dataset. The columns indicate three different types of GAMs: a GAM based on gradient boosting and tailored neural networks \citep{kraus2024interpretable}, a spline-based GAM \citep{hastie1987generalized}, and a tree-based GAM \citep{lou2012intelligible}. Each row highlights one of the five metrics.
These metrics naturally show strong correlations, since all metrics are ultimately designed to correlate with graph complexity. However, our correlation analysis suggests that while the metrics are correlated, they are not redundant, and significant differences emerge for individual graphs.\footnote{The complete correlation analysis is available in a Jupyter notebook in our supplementary materials: \url{https://osf.io/hfjye?view_only=bdb6b9d5d7994fa79ad0fd9826b44e61}}

\subsection{User Study}
\label{sec:user_study}

%\label{subsec:flow}
\textbf{Survey Flow}. Our user study unfolds in three progressive phases, each designed to build upon the familiarization and understanding developed in the previous one.
The initial phase is a screening process using the standardized and widely used Visualization Literacy Assessment Test (VLAT) to assess participants' basic proficiency in graph interpretation \citep{lee2016vlat}. As our study focuses solely on line graphs, we adapt the VLAT to include only the five questions relevant to this type of visualization. To ensure that participants have a deep understanding, we set our qualifying threshold at a 100\% score on our adapted test.

Following the assessment, participants engage in graph value reading tasks, interpreting specific data points within various plots. This phase demands heightened attention to detail and ensures participants are well-versed in extracting key information from the plots. Specifically, we introduce counterfactual-simulation tasks to enhance understanding of the shape plots \citep{doshi2017towards}. Based on the shape plot explanation in Section~\ref{sec:gams}, a related counterfactual-simulation question could be: \textit{Given the input x = 26, under the condition of changing the input value as much as possible: what is the new input value that leads to the output value 10?} Through these tasks, participants are exposed to shape plots with varying visual properties.

Building on the previous task, participants then evaluate their perceived cognitive load imposed by various graphs through activities that include i) ranking graphs by perceived cognitive load, ii) comparing two graphs (i.e., binary choices) deciding which one imposes greater cognitive load, and iii) rating the cognitive load of individual graphs using the 9-point scale developed by \citet{paas1992training}.

%\label{subsec:variables}
\textbf{Measured Variables.} In the empirical evaluation of the visual property metrics, we investigate several key variables to understand the cognitive load associated with shape plots from different types of GAMs:

\textit{Rankings.} We collect ordinal data through participants' ranking of shape plots. By asking participants to order graphs by perceived cognitive load, we obtain a rank-order variable that captures relative cognitive load.

\textit{Binary Choices.} To gauge immediate perceptions of cognitive load, we utilize a binary measure, as suggested by \citet{doshi2017towards}. Participants indicate which graph from a random pair comparison is associated with more cognitive load.

\textit{Cognitive Load Ratings.} Cognitive load can be directly measured using various methods, including subjective rating scales, physiological measures (e.g., eye-tracking, pupillometry), neurological measures (e.g., EEG),  and performance measures (e.g., response time, accuracy) \citep{paas2003cognitive, antonenko2010using}. In our study, we follow \citet{abdul2020cogam} and employ the 9-point scale developed by \citet{paas1992training}, ranging from 1 ("Very, very low mental effort") to 9 ("Very, very high mental effort") to assess participants' self-reported perceived cognitive load when working with shape plots of varying complexity.

These variables are measured directly through participant responses in the survey. The ranking and binary choice variables provide insights into the relative cognitive load imposed by the graphs, while the cognitive load ratings capture a more precise level of perceived cognitive load. Together, they offer a comprehensive view of the cognitive challenges posed by different GAM shape plots. This cognitive challenge can impact their interpretability, as outlined in Section \ref{sec:vp_gc_cl}.

%\label{subsec:participants}
\textbf{Participants.} Overall, we recruit 83 participants using the online platform Prolific. Due to failed VLAT tests, we exclude 26 participants. The average age of participants is 32.21 years, and Table \ref{tab:education_by_gender} provides a demographic summary of the final participants $(N = 57)$.

\begin{table}[h!]
  \centering
  \caption{Summary of participant demographics, stratified by level of education.}
  \label{tab:education_by_gender}
  {\footnotesize
  \begin{tabular*}{\linewidth}{@{\extracolsep{\fill}}
        l 
        S[table-format=2.2] 
        S[table-format=2.2] 
        S[table-format=2.2]
    }
    \toprule
    {Education level} & 
    {\begin{minipage}[t]{0.9cm} % Apply minipage consistently
    \centering
    {Female (\%)} 
    \end{minipage}}
    & {\begin{minipage}[t]{0.9cm} 
    \centering
    {Male (\%)}
    \end{minipage}}
    & {\begin{minipage}[t]{0.9cm} 
    \centering
    {Other (\%)}
    \end{minipage}} \\
    \midrule
    High school diploma               & 10.71 & 37.04 & {--}   \\
    Trade/Tech school certificate     & 14.29 & 3.70  & {--}   \\
    Some college, no degree           & 14.29 & 7.41  & {--}   \\
    Associate's degree                & {--}  & 3.70  & {--}   \\
    Bachelor's degree                 & 42.86 & 33.33 & 50.00  \\
    Master's degree                   & 14.29 & 11.11 & {--}   \\
    Doctorate or PhD                  & 3.57  & 3.70  & 50.00  \\
    \midrule
    \% of total participants          & 47.37 & 49.12 & 3.51 \\
    \bottomrule
  \end{tabular*}
  }
\end{table}

\subsection{Analysis Methods}
\label{sec:analysis}

In this section, we describe our analysis methods used to analyze user measurements and their correlation with different visual property metrics. Our empirical evaluation aims to establish the coherence between measurements based on our metrics and those derived from users' assessment of cognitive load. This analysis is divided into three parts: metric-based models, logarithmic transformation, and validation of metric-based models.

\textbf{Metric-based Models.} To create simple models that describe the relationship between our proposed visual property metrics and perceived cognitive load, we use linear regression. For each metric, we fit a separate linear model and examine how well it matches users' cognitive load ratings, which are obtained by applying the 9-point scale developed by \citet{paas1992training}, as described in Section \ref{sec:user_study}. Each of the 144 plots is rated by at least 12 different participants, and the average of these ratings gives us a perceived cognitive load rating for each plot on a continuous scale from 1 to 9. We consider this average rating to be the most representative indicator of user perception. Our approach allows us to create simple metric-based models that predict the perceived cognitive load of any graph based on its visual properties. As our main evaluation measure, we use $R^{2}$ (coefficient of determination) and also report $MAE$ (mean absolute error), $MSE$ (mean squared error), and the regression $p$-value for each model.

\begin{figure}[h!]
\centering
\includegraphics[width=0.75\linewidth]{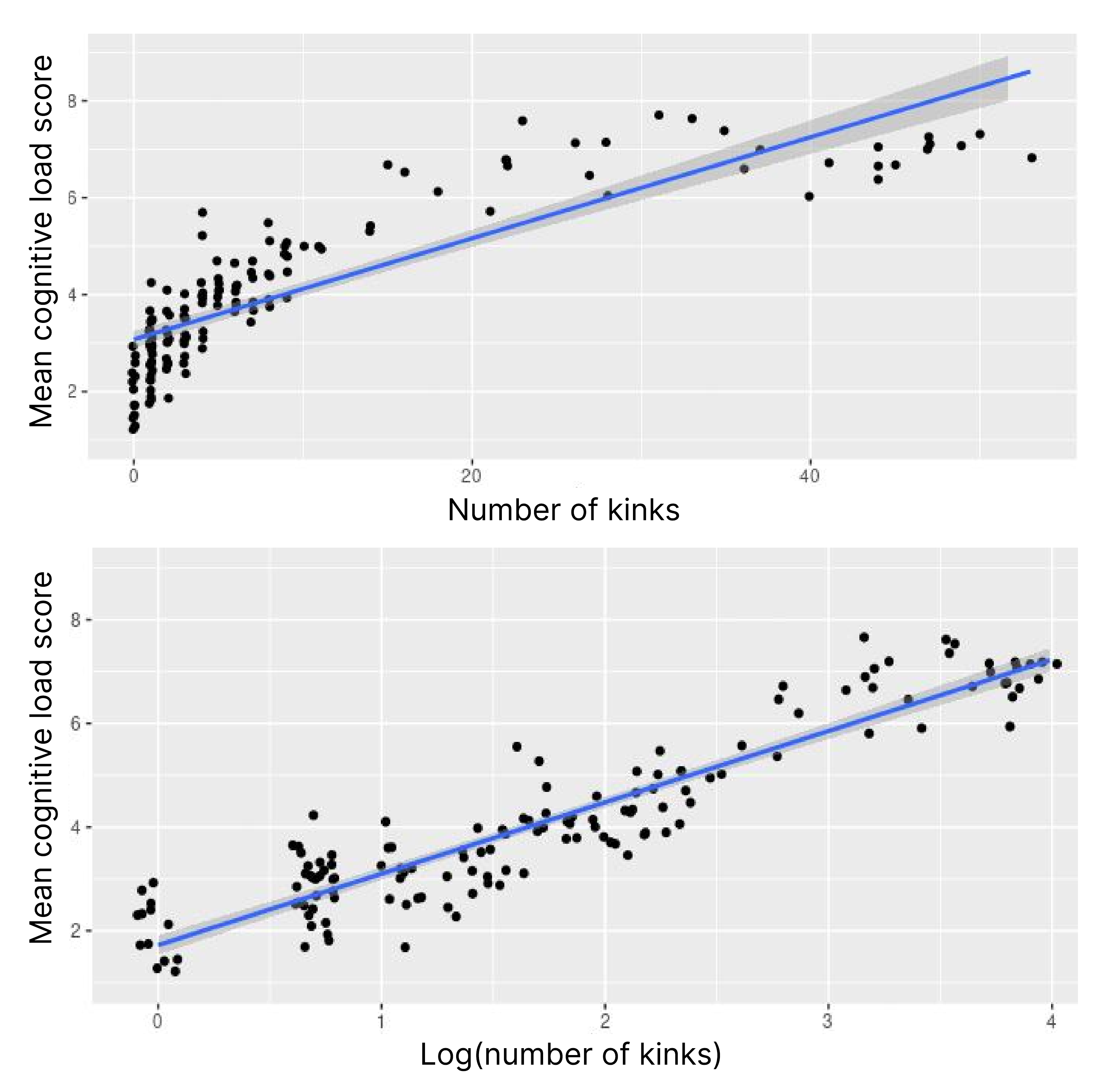}
\caption{Visualization of the logarithmic transformation's effect on the metric-based model for the number of kinks metric.}
\label{fig:log_transform}
\end{figure}

\textbf{Logarithmic Transformation.} As can be seen in Figure \ref{fig:log_transform}, heteroscedasticity is detected in the metric-based models. For plots of mid-range metric values, the models underestimate cognitive load. As a solution, the following logarithmic transformation is applied:

%\vspace{-0.4 cm}
\begin{equation}
\text{log}(M) = \ln(1 + M),
\end{equation}

\noindent
where $M$ is the metric to analyze. This results in a more consistent error distribution and much better model fits. The effect of such a transformation is visually illustrated for the number of kinks metric in Figure \ref{fig:log_transform}.

\textbf{Validation of Metric-based Models.} To further evaluate the quality of our metric-based models, we use the rankings and binary choices of the participants. For each metric-based model, we assess how well the model's predicted cognitive load ratings explains user decisions in terms of rankings and binary choices.

\textit{Rankings.} With 57 participants completing four ranking tasks each, a total of 228 different rankings are obtained. To measure the agreement between our metric-based models and users' rankings, we use the Spearman rank correlation coefficient, denoted as $\rho$. We compute this coefficient for all 228 rankings and report its mean and standard deviation for each metric separately. This approach allows us to evaluate how well our metric-based models align with users' perceived cognitive load while comparing and ranking multiple graphs at the same time.

\textit{Binary Choices.} We extend our analysis to the binary choices. Each participant is presented with eight binary choices between two graphs, resulting in 456 data points. We use these data points to analyze the accuracy and error rate of our metric-based models. In this way, we are essentially testing how often the predicted cognitive load of our models would suggest the same choices as the users. Additionally, we note the frequency of ties, which occur when the metrics for both graphs in a binary choice take the same value.

\textit{Baseline.} To contextualize the quality of our metric-based models in explaining rankings and binary choices, we introduce a baseline. This baseline is derived from the mean cognitive load rating across participants, so that each graph is associated with a numerical value. This numerical value represents users' overall perceived cognitive load associated with that graph and serves as a suitable baseline as it is not grounded on a learned function, but directly on users' perception. Both numerical values -- one resulting from our metric-based-model, the other from the baseline -- are used to test how often they match users' actual decisions in terms of binary choices or rankings as explained above. This baseline highlights that even user-generated ratings have errors in explaining rankings or binary choices due to the inherent ambiguity of human perception. By comparing the performance of our metric-based models to this user-based baseline, we can better assess the effectiveness of our models in capturing human perception of cognitive load. 

\section{Results}
\label{chap:results}
This section presents the effectiveness of our metric-based models in explaining user perceptions of cognitive load. To ensure an accurate assessment, we use a variety of variables. We fit linear models using numerical cognitive load ratings and evaluate their fit, while also using rankings and binary choices to validate the models. %The results of these analyses are detailed in the following subsections.

\subsection{Metric-based Models}
\label{sec:cogload_scores}

We first examine the relationship between visual property metrics and cognitive load ratings (averaged over participants) using linear models and evaluate how well these metric-based models capture user perception by measuring the models' goodness of fit.

\begin{table}[ht]
\centering
\sisetup{print-zero-integer=true}
\caption{Results of the metric-based model analysis.}
\label{tab:cogload}
{\footnotesize
\begin{tabular*}{\linewidth}{@{\extracolsep{\fill}}
  l
  S[table-format=.3]
  S[table-format=1.3]
  S[table-format=1.3]
  S[table-format=<1.3]
}
  \toprule
  {Model} & {$R^{2}$} & {$MAE$} & {$MSE$} & {$p$-value} \\
  \midrule
  Graph length & 0.612 & 0.758 & 1.066 & <0.001 \\
  Polynomial degree & 0.575 & 0.788 & 1.169 & <0.001 \\
  Visual chunks & 0.691 & 0.729 & 0.849 & <0.001 \\
  Number of kinks & 0.732 & 0.692 & 0.737 & <0.001\\
  Avg. kink distance & 0.484 & 0.878 & 1.419 & <0.001 \\
  \bottomrule
\end{tabular*}
}
\end{table}

In Table \ref{tab:cogload}, the graph length, average kink distance, and polynomial degree metric-based models account for 61.2\%, 48.4\%, and 57.5\% of the variance in perceived cognitive load, respectively, with associated $MAE$s of 0.758, 0.878, and 0.788. Notably, the visual chunks model accounts for 69.1\% of the variance with an $MAE$ of 0.729. Remarkably, the number of kinks model outperforms the others, explaining 73.2\% of the variance and having the lowest prediction error (0.692 $MAE$).

%Although each metric-based model shows some explanatory utility, the linear models show indications of heteroscedasticity, resulting in a systematic error in predicting users' cognitive load ratings. For this reason, we propose utilizing a logarithmic transformation of the visual property metrics, as described in Section \ref{sec:analysis}.

\subsection{Logarithmic Transformation}
\label{sec:log_transformation}

Recognizing that the relationships in our data may not fully conform to linear constraints, this section explores the impact of applying logarithmic transformations to our visual property metrics, as detailed in Section~\ref{sec:analysis}. Such an approach allows us to model nonlinear relationships on the original scale while maintaining a linear model in terms of the parameters.

\begin{table}[ht]
\centering
\sisetup{print-zero-integer=true}
\caption{Results of the metric-based model analysis for logarithmic metrics.}
\label{tab:log_metrics}
{\footnotesize
\begin{tabular*}{\linewidth}{@{\extracolsep{\fill}}
  l
  S[table-format=.3]
  S[table-format=1.3]
  S[table-format=1.3]
  S[table-format=<1.3]
}
  \toprule
  {Model} & {$R^{2}$} & {$MAE$} & {$MSE$} & {$p$-value} \\
  \midrule
  {Log(graph length)} & 0.694 & 0.673 & 0.842 & <0.001 \\
  {Log(polynomial degree)} & 0.735 & 0.573 & 0.729 & <0.001 \\
  {Log(visual chunks)} & 0.857 & 0.488 & 0.394 & <0.001 \\
  {Log(number of kinks)} & 0.864 & 0.478 & 0.373 & <0.001 \\
  {Log(avg. kink distance)} & 0.695 & 0.674 & 0.837 & <0.001 \\
  \bottomrule
\end{tabular*}
}
\end{table}

Applying a logarithmic transformation enhances the predictive accuracy of metric-based models by accounting for nonlinear relationships. This leads to improvements in model fit and reduction in prediction errors. Specifically, the $R^{2}$ values increase between 8.2\% and 21.2\%, the $MAE$ decreases in the range from 0.085 to 0.241, and the $MSE$ reduces between 0.364 and 0.582.

The logarithmic transformation helps the metric-based models capture the diminishing effect of incremental increases in a metric at higher values. For instance, the increase in perceived cognitive load when the number of kinks increase from one to two might be greater than the increase when the number of kinks increase from eleven to twelve. %The improvements observed in all metrics after applying the logarithmic transformation illustrate this diminishing effect of increasing visual property metrics on perceived cognitive load.

\subsection{Validation of Metric-based Models}
\label{sec:rankings}

Next, we evaluate the effectiveness of the metric-based models in predicting user rankings and binary choices in terms of cognitive load. %Our goal is to determine which metric-based models best matches user-generated results.

\textbf{Rankings.} To assess the correlation between each metric-based model and user rankings, we focus on the Spearman rank correlation coefficient ($\rho$). A mean $\rho$ of 1 would imply perfect alignment between the metric-based model and all 228 user rankings, which is unrealistic due to the subjective nature of user perceptions. %Therefore, we use our baseline rating derived from other users' cognitive load ratings on a different set of graphs. 
Therefore, we report the baseline derived from mean users' cognitive load (see Section~\ref{sec:analysis}). A model that approaches this baseline indicates a high level of effectiveness in anticipating user perceptions of cognitive load.

In Table \ref{tab:ranking}, all metric-based models exhibit a high average positive correlation with user rankings, highlighting their general alignment with user perception. The model that stands out here is based on the number of kinks metric. This model not only has the highest correlation but also the lowest standard deviation. This implies a robust and consistent agreement with user preferences. In contrast, the average kink distance, while still showing a positive average correlation, has the weakest relationship among the metric-based models listed. Its higher standard deviation further suggests less consistency in matching user preference. The other models occupy the middle ground in terms of their correlation strength.

\begin{table}[ht]
\centering
\sisetup{print-zero-integer=true, % No leading zero for numbers less than 1
        table-format=.3}
\caption{Results of the Spearman rank correlation analysis for each metric, along with a baseline derived from mean users' perceived cognitive load (CL).}
\label{tab:ranking}
{\footnotesize
\begin{tabular*}{\linewidth}{@{\extracolsep{\fill}}
  l
  S % Automatically determines format based on provided numbers
  S
}
  \toprule
  Model & {$mean(\rho)$} & {$sd(\rho)$} \\
  \midrule
  Graph length & 0.688 & 0.421 \\
  Polynomial degree & 0.717 & 0.407 \\
  Visual chunks & 0.677 & 0.429 \\
  Number of kinks & 0.755 & 0.341 \\
  Avg. kink distance & 0.608 & 0.435 \\
  \midrule
  Perceived CL & 0.760 & 0.382 \\
  \bottomrule
\end{tabular*}
}
\end{table}

\textbf{Binary Choices.} Furthermore, we assess the predictive accuracy of our metric-based models in the context of binary choices, where users decide between pairs of plots based on perceived cognitive load. Essentially, we evaluate if the metric-based models would make the same choices as the users. Note that while users were forced to decide, the metric-based models predict the same cognitive load for graphs with identical metric values, leaving the model indecisive in such cases.

Table \ref{tab:binary} presents the accuracy, error rate, and tie frequency for each metric-based model. The results vary, with the visual chunks model showing the lowest accuracy (70.6\%) and the graph length model the highest (78.3\%). The number of kinks model demonstrates strong performance with a high accuracy of 75.4\% and the lowest error rate of 15.6\%, approaching the baseline derived from mean users' cognitive load, which achieves a higher accuracy of 80.3\% but also a higher error rate of 17.5\%. In context, the number of kinks model suggests the same choice as the users 75.4\% of the time and makes the opposite decision 15.6\% of the time. Further, the model is unable to make a choice 9.0\% of the time because the number of kinks is the same in both graphs.

\begin{table}[ht]
\centering
\sisetup{
  print-zero-integer=true, % Don't add leading zero for numbers less than 1
  table-format=.3        
}
\caption{Results of the metric-based models for binary choices for each metric, along with a baseline derived from mean users' cognitive load (CL).}
\label{tab:binary}
{\footnotesize
\begin{tabular*}{\linewidth}{@{\extracolsep{\fill}}
  lSSS
}
  \toprule
Model & {$Accuracy$} & {$Error\ rate$} & {$Tie\ freq.$} \\ 
  \midrule
Graph length & 0.783 & 0.217 &  0.000\\ 
Polynomial degree & 0.739 & 0.178 & 0.083 \\ 
Visual chunks & 0.706 & 0.237 & 0.057 \\ 
Number of kinks & 0.754 & 0.156 & 0.090 \\ 
Avg. kink distance & 0.713 & 0.191 & 0.096 \\ 
  \midrule
Perceived CL & 0.803 & 0.175 & 0.022 \\ 
  \bottomrule
\end{tabular*}
}
\end{table}

\section{Discussion}
\label{chap:discussion}
% interpretation of results

Our study advances ML interpretability by introducing and validating quantitative visual property metrics to assess the cognitive load induced by GAM shape plots. Through a user study with diverse GAM visualizations, we collect human-based measures of perceived cognitive load and evaluate the effectiveness of metric-based models in predicting these measures. While all five metric-based models show predictive utility, the number of kinks metric is the most accurate, explaining 86.4\% of the variance and approaching the user-generated baseline in explaining rankings and binary choices. Based on our results, we propose the following equation to quantify and predict perceived cognitive load ($PCL$):
\begin{equation}
    PCL = 1.724 + 1.377 \times \ln(\text{1 + number of kinks})
\end{equation}

To be put in context, this model can predict cognitive load for the three graphs in Figure \ref{fig:plot-overview}. From left to right, the graphs have 6, 4, and 50 kinks, resulting in cognitive load predictions of 4.40, 3.94, and 7.14, respectively. These predictions suggest that the first and second graphs are similar in users' perceived cognitive load, so a selection between both GAMs should be based on performance. However, the graph of the third GAM may be too complex for the user to work with.
This can also be seen in a preliminary analysis, in which we find that the complexity predicted by our model correlates strongly with the number of participants who make errors when reading values from a graph.\footnote{A detailed analysis of the correlation between predicted complexity and participant errors is available in a Jupyter notebook in our supplementary
materials: \url{https://osf.io/3fe42?view_only=bdb6b9d5d7994fa79ad0fd9826b44e61}}

\textbf{Limitations and Future Research.}
Our study has some limitations offering opportunities for future work. 

First, we focused exclusively on GAMs and the effect of numerical continuous features, without investigating categorical features, more complex ML models or dataset-specific effects. We made this deliberate choice because GAMs are among the best-performing interpretable models \citep{lou2012intelligible, zschech2022gam}, and shape plots accurately visualize their decision logic for numerical features in 2D graphs. Also, there are significant differences between plots produced by various GAM architectures and hyperparameter configurations for numerical features, while categorical features yield more consistent bar plots across GAMs. 

Future work could extend our approach to other model agnostic visualization-based interpretability tools, such as SHAP or LIME. For example, SHAP scatter plots could be evaluated using specialized metrics, such as measuring the area covered by data points or the average range of SHAP values across feature intervals. While this could provide measures of model interpretability, challenges remain, such as the large discrepancies between visualizations and actual model behavior for advanced models, or the inability to read exact values from such scatter plots as participants were asked to do in our experiments. By addressing these challenges, future research could help standardize interpretability metrics across models and features, thereby broadening the applicability of our approach. 

Second, our approach does not provide a comprehensive assessment of interpretability, as the relationship between visual properties and interpretability is complex and can vary based on factors such as user expertise, domain knowledge, perceptual processes and individual working memory \citep{huang2009measuring, forster2023user, shah2002review, peebles2003modeling}. Additionally, the generalizability of our findings may be limited by the specific characteristics of our user sample, as the factors just mentioned could influence the link between visual properties and perceived cognitive load. Further, our study focused on single plots without application context, not accounting for the cognitive load of interpreting multiple features simultaneously or the interaction effects between visualizations and domain- or dataset-specific particularities. Future research could explore extending our metrics to incorporate the complexity of interpreting multiple features concurrently in real-world use cases within certain contexts, bringing us closer to a more generalizable assessment of GAMs' interpretability.

Third, while cognitive load is an important aspect of interpretability, it is not a direct measure of the construct. Interpretability is a complex and multifaceted concept that lacks a clear definition and validated measures \citep{lipton2018mythos, doshi2017towards}. Our study focuses on one specific aspect of interpretability, namely the cognitive load induced by visual properties of GAM shape plots. Future research should develop more comprehensive methods to assess interpretability directly and distinguish it from related constructs.

Fourth, our metric-based models consider one visual property at a time. Although we tested multivariate models, the strong correlations between visual properties limit their effectiveness. These correlations are expected, since all metrics are ultimately designed to correlate with graph complexity. However, our correlation analysis suggests that while the metrics are correlated, they are not redundant, and significant differences emerge, as detailed in our supplementary materials. Regularized linear models show promise in accounting for these correlations, but they require more data to produce robust results.

Additionally, our use of self-reported measures to assess cognitive load has limitations. Although self-reports are widely used due to their practicality \citep{paas2003cognitive}, they may be subject to biases. We mitigated these limitations through a careful study design, but future work could explore objective measures like EEG to directly assess cognitive load \citep{antonenko2010using}. However, such methods require specialized resources and controlled settings, which may limit their feasibility in some research contexts.

\textbf{Implications for Research.}
Our work contributes to research by demonstrating the feasibility of quantifying the cognitive load of GAM shape plots using visual properties. The metrics we develop enable standardized comparisons between GAM shape plots, providing a valuable tool for studying interpretability. Our research bridges cognitive psychology and ML, enriching the theoretical understanding of interpretability with insights from human cognition and perception. The generally good fit of our models to human-perceived cognitive load aligns with principles of cognitive psychology and data visualization. However, the superiority of the number of kinks metric challenges the focus on visual chunks based on trend reversals in the graph cognition literature \citep{abdul2020cogam, carswell1993stimulus}, suggesting that the number of kinks may provide a clearer, more consistent measure of cognitive load in line graphs. We also provide access to a public dataset of shape plots with user-rated cognitive load, facilitating future research on GAM shape plot interpretability.

\textbf{Implications for Practice.}
A practical contribution of our research are the Python implementations that allow for quick quantification of visual properties and prediction of perceived cognitive load, without the need for costly user involvement. Our research lays the foundation for optimizing GAMs for better interpretability at early stages. Just as loss functions guide model performance optimization, our cognitive load models could be a valuable tool to improve interpretability during the model development process. For instance, a model developer could use our Python implementations to assess the cognitive load of GAM shape plots generated during training and refine the model's architecture or hyperparameters to minimize cognitive load while maintaining performance. %As the demand for interpretable AI systems grows across industries, our work has the potential to make a lasting impact by enabling the development of more transparent and user-friendly models that can be readily understood and validated by decision-makers and stakeholders.

\printbibliography

\end{document}